\documentclass[aps,prb, twocolumn,floats,showpacs]{revtex4-1}
\usepackage{graphicx}
\usepackage{color}
\usepackage{bm}

\newcommand{\half}{\mbox{\small $\frac{1}{2}$}}
\begin{document}
\title{ Multiple  Andreev reflections in $s$-wave superconductor-quantum dot- topological superconductor tunnel junctions and Majorana bound states  }
\author{ Anatoly Golub}
\affiliation{ Department of Physics, Ben-Gurion University, Beer Sheva 84105 Israel\\   }
\begin{abstract}
We calculate the current as a function of applied voltage in non-topological s-wave superconductor
 -quantum dot-topological superconductor tunnel junction. We consider the type of TS which hosts two Majorana bound states (MBS) at the ends of a semiconductor quantum wire or of a  chain of magnetic atoms in the proximity with s-wave superconductor. We find that the $I-V$ characteristic of such system in the regime of big voltages has  a typical  two dot shape and  is ornamented by peaks of multiple Andreev reflections. We also consider the other options when the
  zero energy states are created by disorder (here by Shiba states) or by  Andreev zero energy bound states at the surface of quantum dot and superconductor. The later are obtained by tuning the magnetic field to a specific value. Unlike the last two cases the MBS $I-V$ curves are robust to change the magnetic field. Therefore, the magnetic field dependence of the tunneling current can serve as a unique signature for the presence of a MBS.
\end{abstract}
\maketitle
\section {Introduction}
 In recent years the exotic Majorana bound state (MBS) has been the focus of investigations in condensed matter physics. Different platforms for obtaining a MBS and variety of setups for experimental observation were suggested \cite{kitaev,fu,fu2,das,das2,oreg,alicea,nagaosa,aliceaR}. In particular a zero bias peak in the conductance was predicted \cite{been,tanaka,law,flensberg}. Recently  \cite{yazdani} Majorana  fermions where observed at the edge of a topological superconductor (TS) which  was formed by ferromagnetic chain placed in proximity to an  $s$-wave superconductor with strong spin-orbital interaction. The other of the  leading candidates is  semiconductor quantum wire in proximity to an  $s$-wave superconductor - a system  that generates a TS with two MBS's at its ends. A signature of a MBS in such a system
has been detected in tunneling data  in normal metal - TS junctions \cite{kouw,das3,ando}, though the evidence is not conclusive \cite{lee}.

 A setup has been suggested \cite{ueda} for detecting an Aharonov -Bohm  interference between MBS and a quantum dot, predicting structure in the tunneling data. Furthermore, zero frequency shot noise has been studied \cite{demler,us2,blanter}. However, more evidence
of a MBS is needed.

The modified subgap features as signatures of MBS due to multiple Andreev reflections in a weak link between two topological superconductors was addressed in \cite{houzet}.
It has
been shown theoretically that multiple Andreev reflections (MARs) in a weak link between two topological superconductors(i.e., hosting MBS) could cause
novel subgap structures different from the trivial case
which can also be regarded as signatures of the MBS \cite{houzet,aguado2}.
The other more complicated setup was recently
theoretically investigated in\cite{xu}. There the electronic transport through a junction where a quantum dot (QD)is tunnel coupled on both sides to semiconductor nanowires with strong spin-orbit interaction and
proximity-induced superconductivity is analyzed.

Generally, the tunneling through quantum dots integrated in various tunneling systems has been a subject  of considerable interest \cite{rosenov,flensberg2,baranger,golub,lopez,lutchyn,pascu,lutchyn2}. A possible probe for Majorana fermions was suggested in \cite{rosenov} where two MBS that are coupled to quantum dots which themselves interact  with two normal metal leads, can be uniquely tested  by crossed Andreev reflection. The crossed Andreev reflection itself was proposed early in \cite{been3} as a method to probe nonlocality of a pair of MBSs. A simpler setup with a normal lead connected through one quantum dot  to an  MBS was  analyzed
in \cite{flensberg2}. In this paper the nonlinear conductance  as a function of applied bias and gate voltages was calculated in both  cases of interacting and non-interacting QD. The current peaks where used to read off the parity break of the Majorana system. The more complicated setup with one spinless quantum dot connected to two external normal leads and to the one end of p-wave superconducting nanowire was considered in \cite{baranger}. In this paper the peak value of conductance in a TS and non-topological phases was proposed as method to detect MBSs.

The non-Abelian statistics of Majorana fermion states can be tested with the systems without quantum dots by study the half quantum vortices  in a two-dimensional chiral p-wave superconductor \cite{tewari1} or the fluxsons' interferometry in Josephson junction in a TS \cite{eytan}. An interferometer for Majorana fermions edge states  which occur at interphase between superconductor and magnet  placed in the proximity of a topological insulator, was proposed in \cite{kane3}. In this system the MBS transmission can be probed by charge transport. Separated MBSs in a network of nanowires in topological phase have non-Abelian exchange statistics and were suggested for purposes of quantum computation \cite{been4}. To distinguish the MBS conductance peak from the zero-energy peak due to other effects (such as disorder) the tunneling in the presence of dissipation has been considered in\cite{liu}. Here the resistance of the lead is an important parameter that helps to identify MBS peak conductance as function of temperature.
Yet the other evidence for existence of the nanowire Majorana modes  in a simple tunneling structure is based on the fact that nanowire Majorana modes always come in pairs\cite{sarma4}. Therefore, the hybridization due to finite-length wires leads to the splitting of the zero mode. It was shown\cite{sarma4} that this splitting has oscillatory dependence as function of Zeeman energy or chemical potential.

The interacting quantum dot in the Kondo regime as a tunneling link between normal lead and a MBS located at one end of the TS was considered in works \cite{golub,lopez,lutchyn,pascu}. Unlike the standard  normal-quantum dot-normal (N-QD-N)  tunneling systems, the Kondo effect in N-QD-TS junctions predicts a stronger temperature dependence of conductance at $T>>T_K$ ($T_K$ stands for Kondo temperature) \cite{golub,lopez,pascu}. This fact can by used for identification of MBS. A setup with  two normal leads and one QD connected to the Majorana zero mode of the TS  was proposed to provide experiments which can probe Majorana physics by conductance and shot noise measurements \cite{lutchyn2}, wherein the dot may by in the Kondo regime.

Here we consider a simpler case of a tunnel junction s-wave superconductor-quantum dot- topological superconductor (S-QD-TS) where S stands for  topologically trivial s-wave superconductor and TS hosts one MBS at his tunneling end to the quantum dot.  We study the case of large voltages V ( though $eV<\Delta$) which permits  ignoring constant phase-difference. We use the approximation of non-interacting dot ($U=0$). This is justified if $T_K<\Delta$. In this case the Kondo effect has little impact on transport current. Moreover, we  consider the low temperature regime.

If interaction is small  we assume that the charging energy of the dot is much smaller than $\Delta$ and may be  ignored \cite{berg}.
We also consider  the weak tunneling limit when direct tunneling between superconductors is small and, therefore, multiple Andreev reflections due to these direct tunneling events are negligible in the sub-gap region.

Including the QD change in the situation: The transport current acquires a structure  typical for two dot tunneling processes \cite{kouw3}. However, we show that the contributions which come from MBS of the TS can be easily distinguished  from  a random impurity zero energy states inside the gap of a topologically trivial s-wave superconductor. As an example of  such an impurity we take classical magnetic impurity with spin S (Shiba model \cite{shiba,ru}).  The Shiba resonance
is strongly influenced by the applied magnetic field. The same is true  in other case of Andreev zero energy bound states which we also consider in detail.

The structure of the paper is as follows. In Sec.II we introduce model and present the Hamiltonian of tunneling setup. In Sec.III we consider the case of a TS with Majorana bound states at the ends of the  nanowire. Sections IV and V describe two other models without MBS: the impurity zero mode (Shiba resonance) and Andreev zero energy bound states, respectively. Finally, we conclude in Sec.VI. The techniques we include in Appendix.

\section{ The Hamiltonian} The Hamiltonian of our system consists of the topologically trivial s-wave superconductor lead part $H_L$, the quantum dot $H_d$, and the tunnel couplings $H_T$ Hamiltonian. The geometry is depicted  in Fig. 1. Here $t_R,t_L$ define the tunnel couplings between the MBS and the dot, between the dot and the lead.
\begin{figure}
\begin{center}
\includegraphics [width=0.45 \textwidth ]{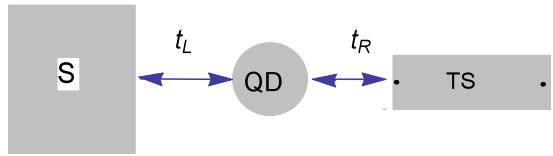}
\caption {Structure of the tunneling junction  which consists of non-topological s-wave superconductor lead, an embedded quantum dot, and a topological superconductor with Majorana fermion at its ends. The  interaction couplings are presented. The  phase $\phi=2eVt$ (we have dropped the constant phase). }
\end{center}
\end{figure}
 $N(0)$ is the density of states of the lead in the normal state and the tunneling widths turn out to be  $\Gamma_{L}=2\pi N(0)t_{L}^2<<t_{L,R}$.
The superconducting s-wave lead is placed at voltage bias $V$ which  is bigger compared to all  other energy scales in the system, including Zeeman energy (though, $V$ is less than the superconducting gap).
 We also assume that the MBS is well separated from other  MBSs, e.g.  at the other end of a TS wire, and therefore neglect the coupling between them.  We write
the Hamiltonian in spin ($s$ matrices) and Nambu (particle- hole space, $\tau$ matrices) as
\begin{eqnarray}
  H_{d}& =& \half d^{\dagger}(\varepsilon s_0\tau_z+ H s_z\tau_0) d\label{Hd}\\
  H_T &=&\half [(t_Lc^{\dagger}(0)   +t_R\gamma \bar{V}^{\dagger}s_0)\times \tau_z d+h.c
  \nonumber
\end{eqnarray}
where $s_0,\tau_0,s_i,\tau_i$ $(i=x,y,z)$ are unit and Pauli matrices, respectively, and $2H$ is the Larmor frequency, including the $g$-factor. The Hamiltonian $H_L$ of the superconducting  lead  has a standard form.
 The lead and dot electron operators are of the form
 $c=(c_{\uparrow},c_{\downarrow},c^{\dagger}_{\downarrow},-c^{\dagger}_{\uparrow})^T$
 and the Majorana fermion operator $\gamma$ comes with the spinor
$\bar{V}_{\varphi}=(e^{i\varphi},e^{i\varphi},
e^{-i\varphi}, -e^{-i\varphi})^T$, the $\varphi$ is the constant phase. The average energy level of the dot is $\varepsilon$.

 Here we use a simplified form of interacting between the MBS and the QD suggested in \cite{flensberg2}. In an other model (model2) which includes only interaction with one spin direction the first and last component of the spinor are replaced by zero (interacting with the down spin of the dot). Model2 may be relevant for strong magnetic fields.

The current operator is defined as $J=e\frac{d}{dt}N_L=-ie[N_L,H]$ and acquires a form $J=(-i/4)j_d$ where
 \begin{eqnarray}
   j_d &=& t_L[ c^{\dagger}(0)\times\tau_0 d-H.c.]
 \end{eqnarray}
 We use the current in Keldysh  space \cite{keldysh,kamenev} ($\hat{j}_d $ ) to construct  the effective action  with source term. In the Keldysh theory the source field consists of two components: the classical $ \alpha_{cl}$ and the quantum  one $\alpha$. The classical part $\alpha_{cl}$ is irrelevant for noise and current calculations and we set it to zero. In this case the source action  has a form
 \begin{equation}\label{source}
A_{sour}=\frac{1}{4} \int_t \alpha \hat{j}_d
\end{equation}

\section{ Majorana bound states at the ends of the topological superconductor}
At first we consider a case with a TS as the right lead. The MBS states exist at  both ends of a topological superconductor. For sufficiently long TS only one MBS is involved in  tunneling. After integrating out the lead and dot operators we arrive at the effective action in terms of Majorana Greens function (GF) which depends on coupling strengths and on quantum source  field $\alpha(t)$
 \begin{eqnarray}
   A_t &=& \frac{1}{2}\int_t \gamma^{T}G_{M}^{-1}\gamma; \,\,\,
   G_{M}^{-1} =G_{M0}^{-1}-\Sigma(\alpha)\nonumber\\
   \Sigma(\alpha)&=&t_R^2 \hat{V}^{\dagger}\tau_3G_d\tau_3 \hat{V}
 \end{eqnarray}
here $G^{R,A}_{M0}(E)=1/(E\pm i\delta)$; the quantum dot GF $G_d(E)=[G_{d0}^{-1}-\Gamma_L g_T]^{-1}$ depends on left lead GF with included source term
  $g_T=T_- gT_+$, where
\begin{equation}\label{T}
    T_{\pm} = \tau_z\times\sigma_0\pm \alpha\tau_0\times\sigma_x/2
\end{equation}
 here $\sigma_{x,y,z}$ are the Pauli matrices in the Keldysh space. In the limit $\alpha\rightarrow0$ we obtain
 \begin{equation}\label{Gd}
    G_d(E)=[G_{d0}^{-1}-\Gamma_L \tau_3 g \tau_3]^{-1}
 \end{equation}
   The GF of the noninteracting dot in magnetic field H has a form
 \begin{equation}\label{Gd}
    G_{d0}^R(E) = [(E+i\delta)s_0\times\tau_0-\epsilon s_0\times\tau_z- H s_z\times\tau_0]^{-1}
 \end{equation}

The Keldysh  GFs of the lead,
\begin{eqnarray}
   g &= &\left(
                                                \begin{array}{cc}
                                                  g^R & g^K \\
                                                  0 & g^A \\
                                                \end{array}
                                              \right)
\end{eqnarray}
in equilibrium   ($ V=0$) $g^R$ has a form
\begin{eqnarray}
  g^R &=& \frac{-i}{2}[a(E)s_0\times\tau_0+b(E)s_0\times\tau_1] \\
  a(E) &=& \frac{|E|\theta(|E|-\Delta)}{\sqrt{E^2-\Delta^2}}+ \frac{E\theta(\Delta-|E|)}{i\sqrt{\Delta^2-E^2}}\\
  b(E)&=&\frac{\Delta}{E}a(E)
\end{eqnarray}
where $\theta(x)$ is a step function equal to one if $x>0$ and is zero otherwise.
The energy gap $\Delta$ describes the lead presented by a topologically trivial s-wave superconductor.
Advanced function (A) is equal to the adjoint of the given  retarded function; and $g^K(E)=(g^R(E)-g^A(E))\tanh(E/2T)$.

The off-diagonal GF of an s-wave superconductor depends on the phase of the order parameter $\exp[\pm i\phi(t)]=\exp[\pm i 2eVt]$.
Therefore, at nonzero voltage V we have a Floquet periodic time dependent problem with a basic frequency of $\omega_0=2eV$. A superconducting lead (topologically trivial)  under  fixed voltage  is described by time dependent GFs. Their Fourier-transforms  are expressed in terms of equilibrium ones  (a generalization  to a 4$\times$4 dimension of the relations from Ref.\cite{arnold} )
\begin{eqnarray}
 g(E,E) &=& g_{11}(E-e V)s_0 P_{+} +g_{22}(E+eV)s_0 P_{-} \nonumber\\
  g(E,E-2 e V) &=& g_{21}(E-e V) s_0 \tau_{+} \nonumber\\
   g(E,E+2 e V) &=& g_{12}(E+e V)s_0\tau_{-} \label{gL}
\end{eqnarray}
where $P_{\pm}=(\tau_0\pm\tau_3)/2$, $\tau_{\pm}=\frac{1}{2}(\tau_x \pm \tau_y)$. The lead GF $g$ may by any function (R,A, or K). We have dropped a constant phase which is justified for not very small voltages. A complete representation of GFs in the Floquet basis is presented in the Appendix.

We evaluate the current  by taking derivatives of the effective action with respect to $\alpha$
 and use dimensionless notations: All energies are taken in units of  $\Delta$.
The total dc current is given by three contributions
\begin{equation}\label{cur}
   j/j_0=\frac{t_R^2\Gamma_L}{2\Delta^3}(j_1+j_2+j_3)
\end{equation}
where $j_0=e/(2\Delta)$ and  $j_1$, $j_2$, $j_3$  are expressed in terms of a Majorana, quantum dot and left lead GFs (see the Appendix).

We calculate the I-V characteristics  of a setup  (Fig1) in the sub-gap region and consider  zero and nonzero magnetic field. It is known
that in low transparency superconductor-normal metal- superconductor (SNS) junctions  the subgap current is small (approaching zero value)\cite{arnold,aguado2}. The tunneling through the dot between  superconducting leads
is responsible for MARs which contribute to the current.
The MBS states, acting as the other dot, however, being structureless  (mixing the spin) are quite robust to the change in magnetic field. Thus we have obtained characteristics (Fig.2) typical for two  dots I-V curves \cite{kouw3}. However, unlike the non-topological case these  I-V curves have different peak positions. In the whole subgap region  current-voltage characteristics weakly depend on magnetic field. This is clearly reflected by Fig.2: The peak position for three values of magnetic field: $H=0$; $0.1\Delta$ and $0.2\Delta$ practically coincide. This is a principal criteria which helps to identify the MBS.

The inset in Fig.2 displays the I-V dependence for model2. Here the current peak is shifted in comparison with the spin mixing model [Eq.1]. The magnetic field dependence shows the bigger shift, although, the peak's height  is more suppressed.  This difference comes out because in  model2  Majorana fermions do not mix the spins.

\begin{figure}
\begin{center}
\includegraphics [width=0.45 \textwidth ]{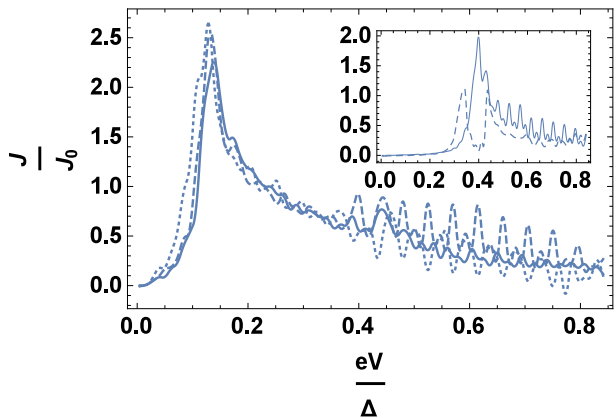}
\caption {Current -voltage characteristics of the tunnel junction (Fig.1). We set the dot energy $\epsilon=- 0.01\Delta$, the temperature $T=0.1\Delta$, and the tunneling widths   $\Gamma_L/\Delta=0.02$ and $t_R^2/\Delta^2 =0.2$ The  lines correspond to zero magnetic field (solid line), to  $H=0.1\Delta$ (dashed line) and $H=0.2\Delta$ (dotted line). inset: the current voltage characteristics in model 2. }
\end{center}
\end{figure}
To calculate the I-V characteristics
the number of Floquet states (2n) is adjusted until the result is insensitive to further increase in n. The calculations include 12 Floquet states.

\section{  Impurity Zero mode in the gap  (no-MBS)}
To prove that we have a clear difference between topological and non-topological cases in this section we  study the current for trivial topology but when, nevertheless,
the zero bound states exist. This may be caused  by Andreev bound states, by localized by
disorder states (impurity), or by surface state as in a d-wave superconductor \cite{us3}. We investigate the I-V characteristics in the case of single Shiba resonance \cite{shiba,ru} when it is tuned to form in-gap zero energy  bound states \cite{glazman, sau}. For a single impurity
in the host superconductor lead (with V=0) the scattering problem can be easily solved \cite{shiba,ru,glazman,sau}. We consider a single (classical) magnetic impurity with spin S at the origin, interacting with the electron states,
\begin{eqnarray}
  H_{imp} &=& -J\vec{S} c_R\bar{s}\times \tau_0 c_R(0) \nonumber
\end{eqnarray}
where $J$ is the exchange strength and $c_R$ stands for the electron operator in the right superconductor. If we define the spin vector as $\vec{S} = S(sin \theta cos \phi , sin \theta sin \phi , cos \theta )$, then at zero order in tunneling  strength $t_R$, Green's function  $ G_{s0}$ of the right lead
 acquires a form (in dimensionless units, and for the frequencies less than the superconducting gap $\Delta$ i.e. $|E|<1$ )
\begin{eqnarray}
  [2G_{s0}^R(E)]^{-1} &=& \frac{E}{\sqrt{1-E^2}} s_0\times\tau_0 +\bar{\alpha}\cos\theta s_z\times\tau_0 - \nonumber\\
  && \frac{1}{\sqrt{1-E^2}}s_0\times\tau_x -H s_z\times\tau_0+\nonumber\\
  &&\bar{\alpha} \cos\phi\sin\theta s_x\times\tau_0+\nonumber\\
 && \bar{\alpha}\sin\phi\sin\theta s_y\times\tau_0 \label{G0}
\end{eqnarray}
where $\bar{\alpha}=\pi N_R JS$ is the dimensionless impurity interaction and $N_R$ is the density of electron states in the right lead. We did not take into consideration the Rashba spin orbit interaction, although, the result for single impurity is  similar to the case without spin-orbit scattering \cite{sau}.  It was shown \cite{glazman} and this can be directly checked by setting to zero the determinant of the matrix (\ref{G0}), that at $\bar{\alpha}\rightarrow 1 $ and $H\rightarrow 0$  we arrive at the zero energy bound states. In the low energy domain close to the in-gap zero mode we can consider $G_{s0}$ at small E. For voltages less than $\Delta$  this level  defines transport. The tunneling interaction with the dot is describe by the same Hamiltonian $H_T$ (\ref{Hd}) where  instead  of $\gamma V^+$  we write projected to low energy domain electron operator$f^{\dagger}$.
As in the case of MBS we integrate out the electron operators of both the left lead  and the quantum dot. Thus we arrive at a general form of the effective action and GF which include interaction with the quantum dot,
\begin{equation}\label{Gi}
    G_s^{-1}=G_{s0}^{-1}-t_R^2 G_d
\end{equation}
The current consists of  three contributions similar to those in Eq.(A15)
however, there is an important difference: The Majorana GF is replaced by the GF of Shiba resonance $ G_s$.
In equilibrium $G_{s0}^R(E)$ (\ref{G0}) is a $4\times4$ matrix in spin and Nambu spaces. In the Floquet basis this matrix has a dimension of
$4(1+2n)\times4(1+2n)$, and the trace (see the Appendix)   operates in this dimension. We calculate the current taking into consideration 12  Floquet states (n=6) using the same set of parameters as in the case of MBS. We consider several values of magnetic field: $H=0$;  $0.1\Delta$;  $0.2\Delta$ and $0.3\Delta$.
In Fig.3 we see a shift in a  peak of transport current as the magnetic field is changed. This does not occur in the MBS case (Eq.1). Unlike the MBS case (Fig.2), here the peak position shifts with  Zeeman energy, and this dependence on H can serve as a possible method to distinguish  the Shiba resonance from the MBS.
\begin{figure}
\begin{center}
\includegraphics [width=0.45 \textwidth ]{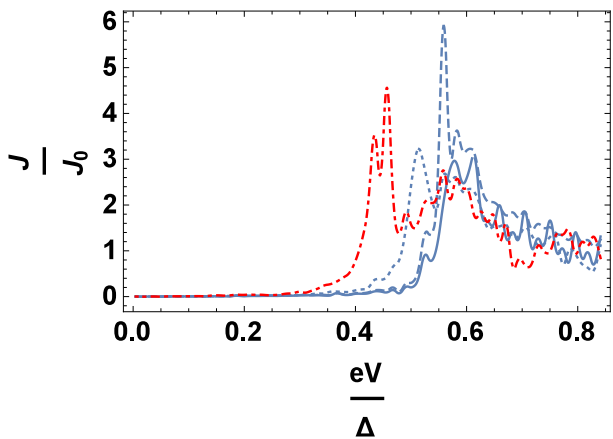}
\caption {(Color online) The same as in Fig2, where, however, the TS is replaced by Shiba resonance at $\bar{\alpha}\rightarrow 1$. For this figure we took direction angles :$\phi=0$,  $\theta=\pi/2$. The other parameters are as in Fig.2 but in addition  dot-dashed curve which corresponds  $H=0.3\Delta$ is included. }
\end{center}
\end{figure}
\begin{figure}
\begin{center}
\includegraphics [width=0.45 \textwidth ]{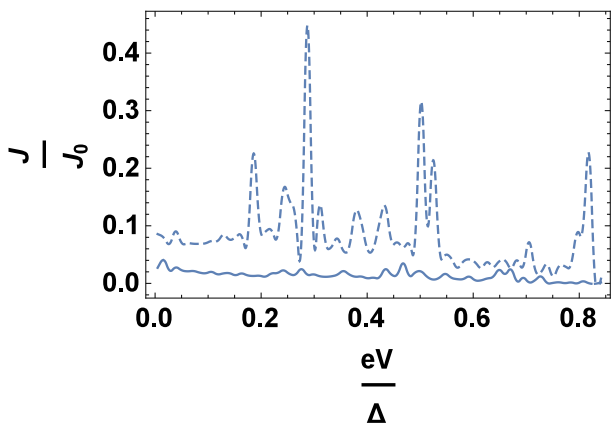}
\caption {(Color online) The tunneling current versus voltage in the case of formation of AZBS at $H=H_0$ (dashed curve) and at H=0 when AZBS are depressed (solid line). Here we have chosen  the dot energy $\epsilon=- 0.4\Delta$, the temperature $T=0.03\Delta$, the tunneling widths   $\Gamma_L/\Delta=0.2$ and $\Gamma_R/\Delta =0.3$  }
\end{center}
\end{figure}

\section{Andreev Zero Bound States}
 Andreev bound states can appear in a system, such as ours when a quantum dot contacts with a superconductor. The zero energy limit mimics the MBS and may be obtained by proper tuning the Zeeman energy. Let us consider setup such as presented by Fig.1 where, however, instead of a topological superconductor on the right hand side we have an s-wave superconductor which is grounded. By  tuning  the magnetic field we intend to get the low energy subspace due to interaction with the s-wave superconductor, i.e., we associate Andreev zero bound states (AZBSs) only with an s-wave superconductor which couples to a quantum dot. Integrating out the electron operators of superconductors (left lead and right) we obtain a total  GF $G_{t}$ of the dot which includes interactions with both superconductors. Actually, $G_t$ has a form of Eq.(6), although, $G_{d0}$ is replaced by $G_{t0}$,
\begin{eqnarray}
     G_{t0}^{-1R}(E)& =& (E+\frac{\Gamma_R E}{2\sqrt{\Delta_R^2-E^2}}+i\delta)s_0\times\tau_0-\epsilon s_0\times\tau_z-\nonumber\\
     && H s_z\times\tau_0-\frac{\Gamma_R\Delta_R}{2\sqrt{\Delta_R^2-E^2}}s_0\tau_x\label{Gt}
\end{eqnarray}
were we, anticipating a low energy domain, consider only the case $|E|<\Delta_R$. It is a direct way to show ( by finding the roots of equation $det[G_{t0}^{-1R}]=0$)  that  the zero energy bound state
can appear when we tune Zeeman energy to the value $H=H_0=\sqrt{\Gamma_R^2/4 +\epsilon^2}$.

 We compute the transport current (see the Appendix) and find the I-V characteristics of the junction (Fig.4). We can clearly distinguish AZBSs which are created at magnetic field $H=H_0$ from the Andreev  bound states created  at  $H\neq H_0$ (here H=0). Many resonances which are shown on Fig.4 correspond to Floquet number shifted by the zero energy pole of the  GF (\ref{Gt}).
Moreover, although, the AZBS can mimic the resonance due to MBS, this resemblance may be destroyed  by a magnetic field different from $H_0$.

\section{ Conclusion} We have applied the standard Keldysh technique \cite{keldysh,kamenev} to evaluate the tunneling current in the setup as presented by Fig.1  As a specific example we  consider a Majorana fermion at the end of a quantum wire which is placed in proximity with a superconductor and under an applied external magnetic field \cite{das,oreg}.
Evidently, control of the magnetic field and the dot-MBS coupling $t_R$ can provide a sensitive test for the MBS detection and may help to distinguish the MBS from other zero bound states \cite {lee} caused either by Andreev bound states or localized by disorder states or by surface states as in d wave superconductors \cite{us3}.
The difficulty with experimental identification of a MBS via the method of a zero bias conduction peak \cite{kouw,das3,ando,lee} is that similar peaks may be due to other low energy bound states \cite {lee2}, such as states localized by disorder \cite{altland}. However, in the experiment \cite{yazdani} a chain of interacting magnetic iron atoms (magnetic dots) on the superconducting lead was investigated. For this system which includes Hubbard interaction in the dot \cite{wen} the theory \cite{altland} is not directly applied.

 We provide the solution of several models: two with the  MBS, the other one is a model in which the MBS is replaced by Shiba  impurity resonance, and the last model represents the AZBS that can appear at the contact of quantum dot and an s-wave superconductor at the specific value of Zeeman energy. We consider multiple Andreev reflections  which are  beyond the small voltage regime.
  We show that for the last two (no MBS) models zero localized states may be identified by strong peak position dependence on the magnetic field. However, in model2 of the MBS that describes the interaction of Majorana fermions with only one spin state of a QD (here spin down)  current peak decreases with magnetic field, although, unlike the model Eq.1 it is shifted. Physically this happens because in the model ( Eq.1) the  interaction with  spins of the  dot involves spin mixing and Majorana fermions act like a Bogolubov quasiparticle, while in model2 spin mixing is excluded.
 A further difficulty with experimental identification is due to the accuracy with which a zero energy state can be determined as function e.g. of a magnetic field \cite{kouw4,das3,ando,lee2}.
Therefore, control of the magnetic field and the dot-MBS coupling $t_R$  provide  an option  for a MBS detection.
\begin{acknowledgments}

 I would like to thank B. Horovitz for stimulating discussions. This research was supported by the ISRAEL SCIENCE FOUNDATION (BIKURA) (grant No. 1302/11).
\end{acknowledgments}

\onecolumngrid
 \appendix

\section {}
\subsection{Greens functions in Floquet space}

Here we obtain the nonequilibrium  GFs of the superconductor (left electrode), the Majorana and the dot Green functions (Eq.4 and Eq.6) as matrices in Floquet space.
 At constant applied voltage V the tunneling between two superconductors is described  by
 GFs which depend on time via the phase of the order parameter \cite{arnold}. The nonequilibrium
 GF of superconductor $\tilde{g}$  acquires a form
\begin{eqnarray}
  \tilde{g}(t,t') &=& \exp[\frac{i\phi(t)\tau_z}{2}] g(t-t')\exp[\frac{-i\phi(t')\tau_z}{2}]
\end{eqnarray}
where $\phi(t)=\phi_0+2eVt$,  $\phi_0$ is a constant phase which we set to zero  and $g(t-t')$ is the equilibrium GF of the superconductor. Due to off-diagonal terms in $g$ the phase exponent does not commute with $g$. Therefore, the Fourier transform of $\tilde{g}$ which depends on two energies, includes energies shifted by a period $2eV$,  actually a multiple of this period (12),
\begin{eqnarray}
  <E|\tilde{g}|E'> &=& \delta(E-E')g(E,E)+\delta(E-E'-2eV)g(E,E-eV)+\delta(E-E'+2eV)g(E,E+eV)
\end{eqnarray}
 The   current is presented as a Fourier series $J(t)=\sum_n\exp[i2eVt] \hat{J}_n(2eV)$. The zero component (n=0) stands for the  averaged current which  includes  integration over E,E'. Therefore, all $\delta$-functions in (A2) are integrated out, and only functions g(E,E') remain.
 If in these GFs we replace E,E' by E+2eVm,E+2eVn then it is convenient to introduce the matrix notation: $g(E+2eVm,E+2eVn)=g_{m,n}(E)$. Because $g$ for every m,n is a $4\times4$ matrix itself we use the indices  $i,k$ to designate the actual matrix element \cite{arnold}.

Let us consider $2N+1$ Floquet states.
As the simplest example we start with the zero order retarded Majorana GF (see Eq.4).
It has no $4\times4$ matrix structure and  thus consists of only diagonal matrix elements  in Floquet space,
\begin{eqnarray}
  G_{M0p,q}^{-1R}(E) &=& \delta_{p,q}(E-2eVN+2eVp)
\end{eqnarray}

Definitions of GFs (Eq.12 ) show that the energy difference between the initial and the final states is the integer multiple of $2eV$. To simplify notations we define $I=integer[\frac{i-1}{4}]; K=integer[\frac{k-1}{4}]$ and $E_K=E-2eV(N-K)$, where $i,k=1,2..., 4(2N+1)$. Thus we have
\begin{eqnarray}
  g_{i,k}(E) &=& g_d [i,k]+ g_+[i,k]+g_-[i,k]\label{g}\\
  g_d[i,k] &=& \delta_{I,K}\{g_{11}(E_K-eV)s_0P_+ +g_{22}(E_K+eV)s_0P_- \}_{i-4K,k-4K}\nonumber\\
  g_+ [i,k] &=& \delta_{I,K-1}g_{21}(E_K-eV)\{s_0\tau_+\}_{i-4(K-1),k-4K}  \nonumber\\ g_-[i,k]&=&\delta_{I,K+1}g_{12}(E_K+eV)\{s_0\tau_-\}_{i-4(K+1),k-4K} \nonumber
\end{eqnarray}
 The matrix structure of the $ 4(2N+1)\times 4(2N+1)$ matrix $g_{i,k}(E)$  consists of $4\times4$ diagonal boxes ($g_d$) and of $4\times 4$ blocks $g_{\pm}$ on each side of the diagonal. The other Keldysh GFs have similar representations. The dot GF (Eq.6 ) includes the lead GF $g$ as its non-equilibrium part, therefore, we can write the total inverse  dot GF in the form, such as $g$ (Eq.A4)
\begin{eqnarray}
  G_{d i,k}^{-1R}(E) &=& G_1 [i,k]+ G_+[i,k]+G_-[i,k]\label{Gi}\\
  G_1[i,k] &=& \delta_{I,K}\{E_Ks_0\tau_0-\epsilon s_0\tau_z-H s_z\tau_0
  -\Gamma_L[g_{11}(E_K-eV)s_0P_+ +g_{22}(E_K+eV)s_0P_-] \}_{i-4K,k-4K}\nonumber\\
  G_+ [i,k] &=& \delta_{I,K-1}\Gamma_Lg_{21}(E_K-eV)\{s_0\tau_+\}_{i-4(K-1),k-4K}  \nonumber\\
  G_-[i,k]&=&\delta_{I,K+1}\Gamma_L g_{12}(E_K+eV)\{s_0\tau_-\}_{i-4(K+1),k-4K}\nonumber
\end{eqnarray}
The dot GF $G_{d i,k}$ is obtained by taking the inverse of Eq.(\ref{Gi}).

Total Majorana GF, although, which depends on dot function  $G_{d }$ (Eq.A4) has no spin and particle-hole presentation. It is a matrix only in Floquet space $(2N+1)\times (2N+1)$.  Using definition of spinor $\hat{V_{\varphi=0}}$ (see Eqs.1 and 4 ) we find
\begin{eqnarray}
  G_{M p,q}^{-1 R}(E) &=& G_{M0 p,q}^{-1R}(E)-\Sigma_{p,q}^R(E)  \label{GM}\\
  \Sigma_{p,q}^R(E)&=&t_R^2\{G_{d 1 + 4p, 1 + 4q}^R(E)+G_{d 2 + 4p, 2 + 4q}^R(E)+G_{d 3 + 4p, 3 + 4q}^R(E)+G_{d 4 + 4p, 4 + 4q}^R(E)+\nonumber\\
  && G_{d 1 + 4p, 3 + 4q}^R(E)-G_{d 2 + 4p, 4 + 4q}^R(E)+G_{d 3 + 4p, 1 + 4q}^R(E)-G_{d 4 + 4p, 2 + 4q}^R(E)-\nonumber\\
  && G_{d 1 + 4p, 4 + 4q}^R(E)+G_{d 2 + 4p, 3 + 4q}^R(E)+G_{d 3 + 4p,2 + 4q}^R(E)-G_{d 4 + 4p, 1 + 4q}^R(E)+\nonumber\\
  && G_{d 1 + 4p, 2 + 4q}^R(E)+G_{d 2 + 4p, 1 + 4q}^R(E)-G_{d 3 + 4p,4 + 4q}^R(E)-G_{d 4 + 4p, 3 + 4q}^R(E) \}
\end{eqnarray}
here $p,q=0,1,2...,2N$. Inverting Eq.(\ref{GM}) we arrive at the effective Majorana GF.

The  inverse GF of the Shiba states in the low energy limit close to the in-gap zero (at $\tilde{\alpha}=1$) replaces the Majorana GF $G_{M0 p,q}(E)$ in the expressions for the tunneling current. The effective  Shiba state Green's function (Eq.15 ) has the self-energy part which is determined by interaction with the dot. In the Floquet basis this GF is $4(2N+1)\times 4(2N+1)$ matrix which has the form
\begin{eqnarray}
  G_{s p,q}^{-1R}(E) &=& G_{s0 i,k}^{-1R}(E)-t_R^2 G_{d i,k}^{R}(E) \\
  G_{s0 i,k}^{-1R}(E) &=& \delta_{I,K}\{E_Ks_0\tau_0-\epsilon s_0\tau_z-Hs_z\tau_z+\cos\theta s_z\tau_0 +\cos\phi\sin\theta s_x\tau_0+\sin\phi\sin\theta s_y\tau_0\}_{i-4K,k-4K}
\end{eqnarray}

\subsection{The tunneling current }
Let us at first consider the tunneling current in the S-QD-TS(MBS) junction.
We evaluate the current  by taking derivatives of the effective action with respect to $\alpha$,
\begin{equation}\label{cur}
   j(t) =\frac{e}{4}Tr\int dt_1\int dt_2 G_M(t_1t_2)(\frac{\delta\Sigma(\alpha t_2t_1)}{\delta\alpha(t)})_{\alpha\rightarrow 0}
\end{equation}
where Tr acts in Keldysh space. Explicitly the derivative acquires the form
\begin{eqnarray}
  \frac{\delta\Sigma(\alpha t_2t_1)}{\delta\alpha(t)} &=& t_R^2\Gamma_L \int dt_3\int dt_4\hat{V^{\dagger}}G_d(t_2t_3)\frac{\delta g_T(t_3t_4)}{\delta\alpha(t)}G_d(t_4t_1)\hat{V}\label{sigma}\\
  (\frac{\delta g_T(t_3t_4)}{\delta\alpha(t)})_{\alpha\rightarrow 0} &=& \frac{1}{2}[g(t_3,t)\delta(t-t_4)\sigma_x \tau_z-\sigma_x\tau_z \delta(t-t_3)g(t,t_4)]\label{dg}
\end{eqnarray}
Performing the trace  in the Keldysh space we obtain  several contributions to the current where, in addition to retarded and advanced GFs, the Keldysh component of the GF is also involved. From Eqs.and 6 we obtain for these GFs,
\begin{eqnarray}
  G_M^K&=&t_R^2G_M^R\hat{V^{\dagger}}\tau_zG_d^K\tau_z\hat{V}G_M^A \label{GMK} \\
   G_d^K &=&\Gamma_L G_d^R\tau_z g^K\tau_zG_d^A\label{GdK}
\end{eqnarray}

We consider the time averaged
 transport current. Only zero  multiple of 2eV  in the Fourier series  contributes to the current. In this case we use the Fourier-transform representation of the GFs (\ref{Gi}) and (\ref{GM}). The current is presented by a trace of proper combinations of these functions in Floquet  space. Inserting  the expressions Eq.(\ref{sigma}) and (\ref{dg}) into Eq.(\ref{cur}), performing the trace in the Keldysh space we arrive at a  final  form of  current in the S-QD-TS(MBS) junction.
 The total dc current is given by three contributions where for the last two we use Green,s functions  Eq.(\ref{GMK}) and (\ref{GdK}),
\begin{equation}\label{current}
   j/j_0=\frac{t_R^2\Gamma_L}{2\Delta^3}(j_1+j_2+j_3)
\end{equation}
where $j_0=e/(2\Delta)$ and  $j_1$, $j_2$,and $j_3$ acquire the form
\begin{eqnarray}
  j_1 &=&tr \int dE  Re[G_M^R\bar{V}^+\tilde{G}_d^Rg^K\tau_z \tilde{G}_d^R\bar{V}]
  \label{j1}
  \end{eqnarray}
  \begin{eqnarray}
  j_2 &=& \frac{\Gamma_L}{\Delta}tr\int dE  Re[G_M^R\bar{V}^+\tilde{G}_d^Rg^K \tilde{G}_d^A(g^A\tau_3-
  \tau_3 g^R)\tilde{G}_d^R\bar{V}]\label{j2}
  \end{eqnarray}
  \begin{eqnarray}
  j_3&=&\frac{\Gamma_L t_R^2}{2\Delta^3}tr\int dE Re[G_M^R\bar{V}^+\tilde{G}_d^Rg^K \tilde{G}_d^A\bar{V}
  G_M^A\bar{V}^+\tilde{G}_d^A(g^A\tau_3-
  \tau_3 g^R)\tilde{G}_d^R\bar{V}]\label{j3}
\end{eqnarray}
Here $\tilde{G}_d^{R,A}=\tau_3 G_d^{R,A} \tau_3 $; $tr$ stands for the trace over the Floquet states, and $G_M^{R,A}$ are the matrices in the Floquet basis of dimension $(1+2N)\times(1+2N)$, the same as the blocks $[\bar{V}^+...\bar{V}]$.

This fact is principal: It distinguishes topological case (with the TS and the MBS) from the trivial normal zero level states inside the gap (here AZBS and Shiba resonance). Indeed the expression for the current in the case of Shiba zero states (i.e., we consider a junction S-QD-S(with the Shiba state)   coincides with Eqs.(\ref{j1})-(\ref{j3}) if: (i) We replace Majorana GFs $G_M$ by GFs of Shiba zero states, (ii) drop spinors $\bar{V}^+ , \bar{V}$; and (iii) take the trace over the space $4(2N+1)\times4(2N+1)$.

We also calculate the current in the case of AZBSs. The transport current through the dot in a setup such as that shown in Fig1 of the main text, is described by Eq.(A10) where instead of $G_M$ and $\Sigma$ we have $G_t=[G_{t0}-\Sigma_g]^{-1}$ and $\Sigma_g =\Gamma_L g_T$, correspondingly, and
\begin{equation}\label{sen}
   \frac{\delta\Sigma_g( t_2t_1)}{\delta\alpha(t)}  =\Gamma_L(\frac{\delta g_T(t_2t_1)}{\delta\alpha(t)})_{\alpha\rightarrow 0}
\end{equation}
With the help of  Eq.16 and Eq.A19 we obtain
\begin{equation}\label{ja}
  j/j_0 =\frac{\Gamma_L}{\Delta}tr \int dE  Re[G_t^R\tau_zg^K(1+\Gamma_L\tilde{G}_t^Ag^A)]
\end{equation}
where $\tilde{G}_t^A=\tau_z G_t^A \tau_z$ and the trace acts in the space $4(2N+1)\times4(2N+1)$.

\def\up{\uparrow}
\def\down{\downarrow}
\def\ud{\uparrow\downarrow}
\def\du{\downarrow\uparrow}
\def\uu{\up\up}
\def\dd{\down\down}
\def\Tr{\mbox{Tr}\,}
\def\Im{\mbox{Im}}
\def\Re{\mbox{Re}}

\setcounter{figure}{0}
\setcounter{equation}{0}
\makeatletter
\renewcommand{\thefigure}{S\@arabic\c@figure}
\renewcommand{\theequation}{S\@arabic\c@equation}
\makeatother


\end{document}